\documentclass[a4paper,prl,twocolumn,superscriptaddress,showpacs]{revtex4}

\usepackage{amssymb,graphicx}

\begin{document}

\title{The geometric structure of the Landau bands}

\author{J.~Br\"uning}
\affiliation{Helmholtz-Zentrum, Humboldt-Universit\"at zu Berlin,
Unter den Linden 6, Berlin 10099 Germany}
\affiliation{Institut f\"ur Mathematik, Humboldt-Universit\"at zu Berlin,
Rudower Chaussee 25, Berlin 12489 Germany}
\author{S.~Yu.~Dobrokhotov}
\affiliation{Institute for Problems in Mechanics, Russian Academy of Sciences,
pr.~Vernadskogo 101, Moscow 117526 Russia}
\author{V.~A.~Geyler}
\affiliation{Laboratory of Mathematical Physics, Mordovian State University,
Saransk 430000 Russia}
\author{K.~V.~Pankrashkin}
\affiliation{Institut f\"ur Mathematik, Humboldt-Universit\"at zu Berlin,
Rudower Chaussee 25, Berlin 12489 Germany}
\affiliation{Institute for Problems in Mechanics, Russian Academy of Sciences,
pr.~Vernadskogo 101, Moscow 117526 Russia}

\begin{abstract}
We have proposed a semiclassical
explanation of the geometric structure of the spectrum
for the two-dimensional Landau Hamiltonian with
a two-periodic electric field without
any additional assumptions on the potential.
Applying an iterative averaging procedure
we approximately, with any degree of accuracy,
separate variables and
describe a given Landau band as the spectrum
of a Harper-like operator. The quantized Reeb graph
for such an operator is used to obtain 
the following structure of the Landau band:
localized states on the band wings and
extended states near the middle of the band.
Our approach also shows that different Landau bands
have different geometric structure.
\end{abstract}


\maketitle

The standard theories of the integer quantum Hall effect
\cite{Lau} are modelled by the Hamiltonian for a charged particle
in a heterostructure with a uniform magnetic field; they are based on the 
following assumption concerning the structure of the spectrum:
the spectrum consists of broadened Landau levels (Landau bands)
generated by  extended states near the middle of each band and by localized 
states in the band wings; then pinning the Fermi level by the localized states
causes the occurrence of plateaus in the conductance curve
(see \cite{PG} for details).
This specific spectral structure is usually related to the presence of 
impurities or other defects in the quantum Hall system~\cite{JP}. 
Indeed, for the Gauss or Lloyd ensembles of impurities the density of states 
has the behavior required for the explanation of the quantum Hall effect 
\cite{Weg}. Moreover,  using ideas from \cite{PF} it can be rigorously shown 
that the desired spectral structure is displayed by the two-dimensional Landau 
Hamiltonian with substitutional point impurities arranged at the
nodes of a lattice, assuming a quasi-periodic distribution of the coupling
constants \cite{GM2}. The same holds for point impurities with
random positions on the system plane and with random distribution
of the coupling constants \cite{AG}.

The aim of the present article is to show that this special spectral structure 
can also be verified by semiclassical approximation for
a purely periodic Hamiltonian with a uniform magnetic field
without any additional assumptions on the potential. Note that a 
semiclassical approach to the quantum Hall effect has been considered in 
\cite{EB} for a Hamiltonian with a disordered potential, within the
framework of percolation theory on the Chalker--Coddington network.
The results of \cite{EB} confirm the intuitive picture which relates
the extended states in the band center with the central drift of an open 
cyclotron orbit, and the localized states in the band wings with closed orbits.
This intuitive picture was rigorously justified in \cite{BD}
using Maslov's canonical operator \cite{Mas}.
Below we offer some consequences of the results from
\cite{BD} which describe in detail the semiclassical spectrum of the
periodic Landau Hamiltonian and show, in particular, the desired spectral 
structure.

We use the Landau gauge for the Hamiltonian of a magneto-Bloch 
electron,
$$
\hat H=\frac{\hbar^2}{2m}\big((-i\partial_1+(eB/c\hbar) x_2)^2 -\partial_2^2\big)
+V(x_1,x_2),
$$
where the potential $V$ is periodic with respect to the lattice $\Lambda$ 
generated by the vectors $\mathbf{a}_1=(L,0)$ and $\mathbf{a}_2$;
we assume that $V$ is a real-analytic function.
Introducing the cyclotron frequency $\omega_c=|eB|/cm$, the magnetic
length $l_M=(\hbar/m\omega_c)^{1/2}$, and passing to dimensionless
potential energy $v=V/\max\,V$ and to coordinates $\mathbf{X}=\mathbf{x}/L$,
we rewrite $\hat H$ in the form
$\hat H=mL^2\omega_c^2\hat H^0$
where
\[
\hat H^0=\frac{1}{2}\left[\left(\hat P_1+X_2\right)^2+\hat P_2^2\right]
+\varepsilon_V v(X_1,X_2)\,.
\]
Here $\hat P_j=-i\varepsilon_B\partial/\partial X_j$ and the
dimensionless quantities $\varepsilon_B=(l/L)^2$ and
$\varepsilon_V=\varepsilon_B\max |V|/\hbar\omega_c$ will be considered
as small parameters in our spectral problem.
Let us estimate these parameters in a typical situation: for a field strength
$B\simeq 10$T the magnetic length is of order 10nm; the characteristic
length $L$ for periodic arrays of quantum dots or antidots in GaAs is
of order 200--500nm, such that $\varepsilon_B\sim 10^{-3}$; and using for 
GaAs the electron effective mass $m=0.067m_e$ we estimate $\hbar\omega_c$
as 15meV, implying that for $V$ with values in the range 1--25 meV \cite{GLD}
we have $\varepsilon_V\lesssim \varepsilon_B$.

The main difficulty of the problem under study consists in the 
non-integrability of the corresponding classical system. At first sight,
the presence of the small parameter $\varepsilon_V$
suggests the use of perturbation theory;
but it is readily seen that this will only be possible if we impose additional 
relations between $\varepsilon_B$ and $\varepsilon_V$, because
$\varepsilon_B$ is also small. Moreover, using perturbation
theory we can obtain only rather rough spectral information. 

The smallness of $\varepsilon_B$, however, will 
allow us to describe the fine structure of each Landau band by 
using a semiclassical approach.

The trajectories of the classical system with the unperturbed Hamiltonian
$\frac{1}{2}\big((P_1+X_2)^2+P_2^2\big)$
on the plane $(X_1,X_2)$ are the cyclotron orbits with radius $\sqrt{2I}$
and angle $\varphi$ centered at a point $(y_1,y_2)$. Therefore, it is
reasonable to introduce new canonical variables: the generalized momenta
$I$, $y_1$ (or $p$, $y_1$) and the positions $\varphi$, $y_2$
(or $q$, $y_2$) according to the formulas
\[
\begin{array}{c}
\displaystyle X_1=q+y_1,\quad P_1=-y_2,\quad X_2=p+y_2, \quad P_2=-q\\
\displaystyle p=\sqrt{2I}\cos\varphi,\quad q=\sqrt{2I}\sin\varphi\,.
\end{array}
\]
In these variables the classical Hamiltonian $H^0$ takes the form
$H^0=I+\varepsilon_V v(\sqrt{2I_1}\sin\varphi+y_1,\sqrt{2I}\cos\varphi+y_2)$.

This representation suggests that we average $H_0$ over the angle $\varphi$, 
and describe the central drift of the cyclotron orbits by means of
the averaged Hamiltonian
\begin{equation}
                 \label{1}
\begin{array}{c}
\displaystyle {\mathcal H}^{\rm av}(I,y_1,y_2;\varepsilon_V)=\frac{1}{2\pi}
\int\limits_0^{2\pi}H^0\,d\varphi\\
\displaystyle{}=I+\varepsilon_VJ_0(\sqrt{-2I\Delta_{\mathbf{y}}})v(y_1,y_2),
\end{array}
\end{equation}
where $J_0$ denotes the Bessel function of order zero; here
the second term on the right-hand side results from the action of the 
pseudodifferential operator
$J_0(\sqrt{-2I\Delta_{\mathbf{y}}})$ on $v$.
Next, it can be shown that there is a canonical change
of variables $(p',q',\mathbf{y}')=(p,q,\mathbf{y})+O(\varepsilon_V)$
such that
\[
H^0(p,q,\mathbf{y};\varepsilon_V)=
{\mathcal H}^{\rm av}\big(\frac{1}{2}(p'^2+q'^2),\mathbf{y}';\varepsilon_V\big)
+O(\varepsilon_V^2),
\]
but the error estimate $O(\varepsilon_V^2)$ is not good enough to
describe the desired fine structure of the spectrum of $H^0$.
The crucial point, then, is the
following assertion which is proved in \cite{BD} by suitably iterating the
averaging procedure:

{\it There is a canonical change of variables
$(P,Q,\mathbf{Y})=(p,q,\mathbf{y})+O(\varepsilon_V)$ such that
in the new coordinates $(P,Q,\mathbf{Y})$ the classical
Hamiltonian system, $H^0$, becomes integrable modulo
$O(\varepsilon_V^\infty)$.
More precisely, $H^0(p,q,\mathbf{y};\varepsilon_V)
={\cal H}^0(\frac{1}{2}(P^2+Q^2),Y_1,Y_2;\varepsilon_V)+
O(\varepsilon_V^\infty)$,
where the first term is real-analytic and all the terms are
two-periodic in $\mathbf{Y}$}.

It is worthwhile to note that the first step of the mentioned
iteration procedure gives just the function ${\mathcal H}^{\rm av}$.

Now we quantize ${\mathcal H}^0$ considering the
pairs $\{P,Q\}$ and $\{Y_1,Y_2\}$
as pairs of canonically conjugated variables, i.e. we  set
$\hat P=-i\varepsilon_B\partial/\partial Q$, $\hat Y_1=-i\varepsilon_B\partial/\partial Y_2$.
By the Correspondence Principle, a canonical transformation in classical mechanics
should correspond to a unitary transformation of the quantum
mechanical state space. In our case
this expectation is expressed in the fact that the spectra of the 
operators $\hat H^0$ and $\hat{\mathcal H}^0$ coincide 
semiclassically, up to $O(\varepsilon_B^\infty+\varepsilon_V^\infty)$.
Now, since $\hat P$ and $\hat Q$ commute with $\hat Y_j$ ($j=1,2$), the quantum
Hamiltonian $\hat{\mathcal H}^0$ commutes with the Hamiltonian of the
harmonic oscillator $\frac12(\hat P^2+\hat Q^2)$.
Hence
the eigenfunctions $\Psi$ of $\hat{\cal H}^0$ may be chosen in the form
$\Psi(Q,Y_2)=\psi_n(Q)\varphi_n(Y_2)$, where $\psi_n$ are the eigenfunctions
of the harmonic oscillator corresponding to the eigenvalues
$E_n=(n+1/2)\varepsilon_B$, $n=0,1,\ldots$, and the functions $\varphi_n$
satisfy the equation
\begin{equation}
                 \label{2}
\hat{\mathcal H}_n \varphi_n =E\varphi_n.
\end{equation}
Here $\hat {\mathcal H}_n$ is a pseudodifferential operator obtained as the 
quantization of the classical Hamiltonian
$\mathcal{H}_n(Y_1,Y_2)={\mathcal H}^0(E_n,Y_1,Y_2,\varepsilon_V)$.
As $mL^2\omega_c^2\varepsilon_B=\hbar\omega_c$, the number
$E_n$ is nothing but the $n$-th Landau level.
Thus the spectrum of each operator $\hat{\mathcal H}_n$
arises from broadening the $n$-th Landau level into a band
under the influence of the periodic potential.
More precisely, in our approach each Landau band is described by the
one-dimensional equation (\ref{2}), which depends on the band,
and the original spectral problem is reduced
to a family of one-dimensional spectral problems.
This reduction makes it now possible to describe
the fine structure of each Landau band using
established methods of semiclassical approximation, since we only 
have to deal with an integrable Hamiltonian.

\begin{figure*}\centering
\includegraphics[width=130mm]{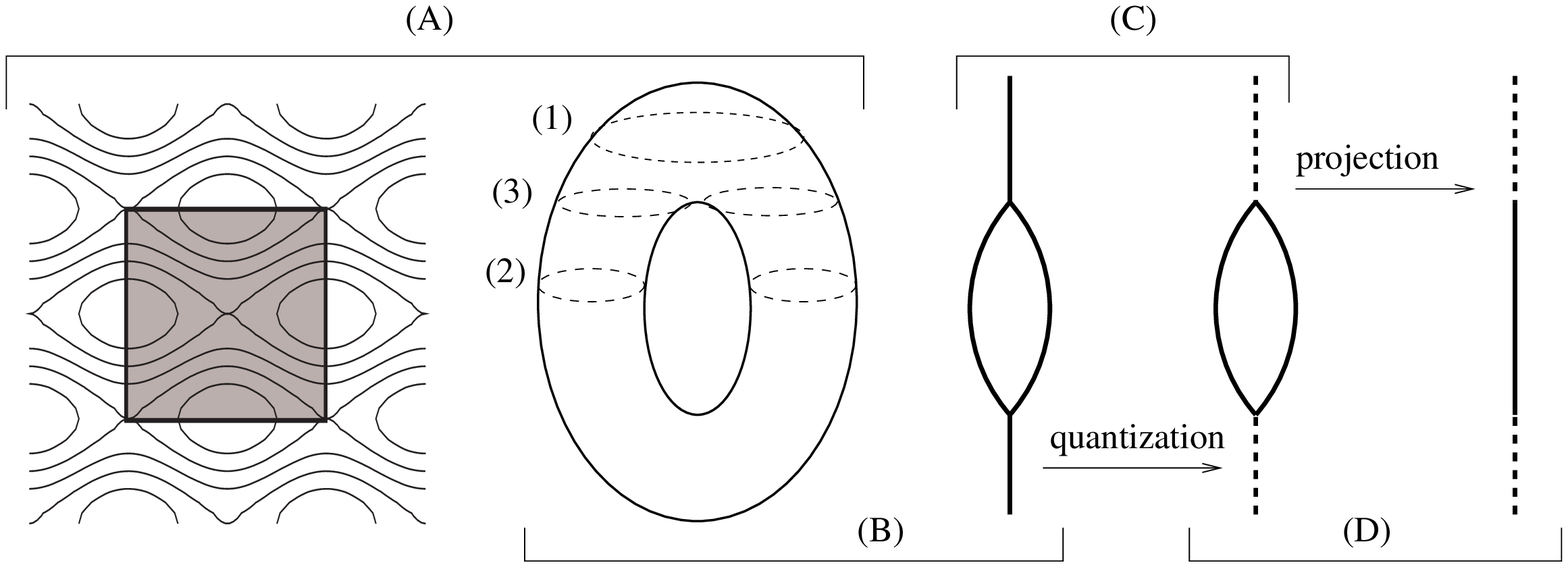}\\[\bigskipamount]
\includegraphics[width=130mm]{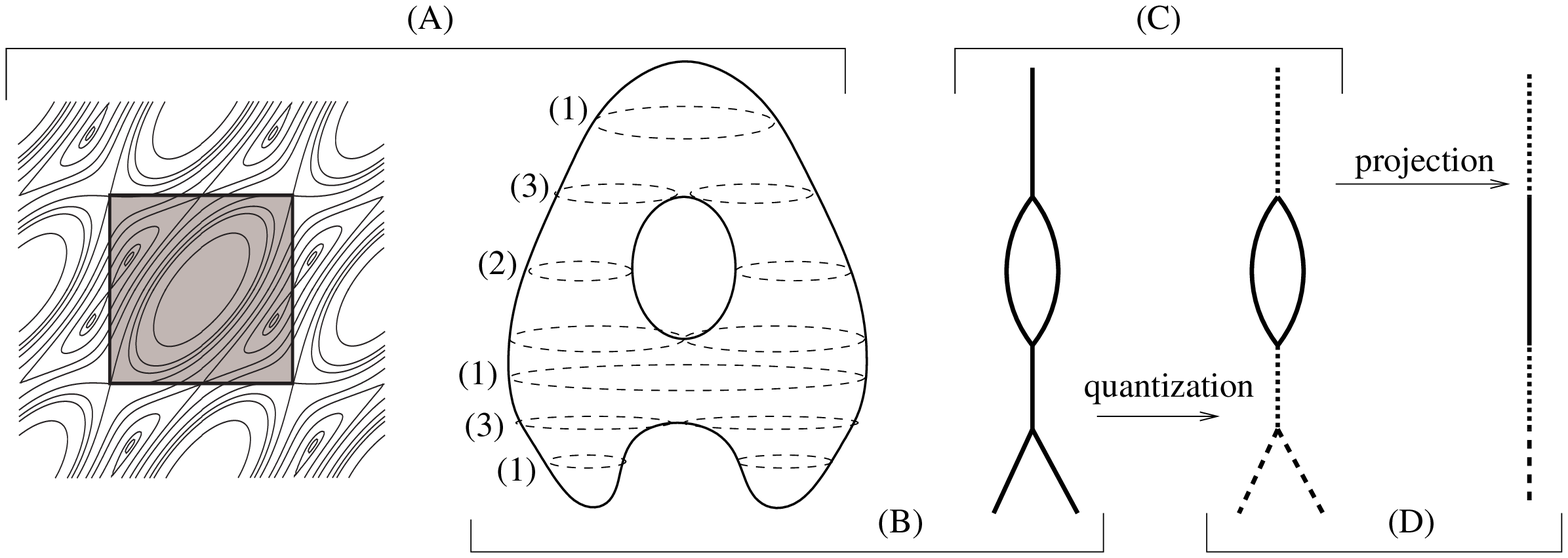}\\
\caption{Two examples of the geometric structure
of the Landau band:
(A) Level curves of a Hamiltonian
on the torus and their realization through the height function;
(B) Trajectories on the torus: (1) contractible closed ones,
(2) non-contractible closed ones, (3) separatrices, and the Reeb graph
of the Hamiltonian; (C) Applying the Bohr-Sommerfeld quantization;
(D) Relationship between the Reeb graph and the Landau band}\label{fig1}
\end{figure*}

In view of the established periodicity,
${\mathcal H}_n$ can be treated as a Hamiltonian on the two-torus; 
therefore, only the following three types of trajectories occur
for the corresponding Hamiltonian system:
(1) contractible closed curves and
extremum points, (2) non-contractible closed curves,
(3) separatrices and saddle points. The concept
of the \emph{Reeb graph}~\cite{Fom}
has proved a very useful tool in describing the topological structure of
the space of trajectories. Namely, points of the Reeb graph
are in a one-to-one correspondence with connected components of
the level sets of the Hamiltonian, in such a way that
separatrices correspond to the branching points of the graph.
If one represents $\mathcal{H}_n$ as the height function
of a suitably deformed torus, then construction of the Reeb graph
becomes especially evident, which is illustrated in Fig.~\ref{fig1}A
and~\ref{fig1}B. .

Obviously, contractible trajectories on the torus
are covered by families of \emph{closed} trajectories on the plane $(Y_1,Y_2)$,
and non-contractible ones are covered by families of
\emph{non-closed} (but periodic) trajectories on this plane.
Hence, to obtain the semiclassical approximation to
the spectrum of $\hat{\mathcal H}_n$,
i.~e., to obtain the semiclassical structure of the
$n$-th Landau band, one has to quantize
only the closed trajectories, using the Bohr-Sommerfeld rule;
thus we obtain the so-called quantized Reeb graph (Fig.~\ref{fig1}C).
Returning to the original coordinates
$(\mathbf{P},\mathbf{X})$ we see that the closed trajectories
of ${\mathcal H}_n$ generate two-dimensional tori. It is well known that
the semiclassical eigenfunctions corresponding to these tori
decrease exponentially outside the classically allowed regions
(i.~e. the projections of the tori onto the $\mathbf{X}$-plane), 
hence correspond to localized states.
On the other hand, open trajectories in the $\mathbf{Y}$-plane correspond
to two-dimensional cylinders in the phase space $(\mathbf{P},\mathbf{X})$.
Since those are non-compact the corresponding states must be extended; 
they can be constructed using Maslov's canonical operator ~\cite{Mas}.

Our considerations show that the $n$th Landau band in the spectrum of $\hat H$ 
is structurally completely  determined
by the Reeb graph of $\mathcal{H}_n$ (see Fig.~\ref{fig1}D).
It is evident from this correspondence that, for a \emph{generic} two-periodic 
potential $V$, the ``highest'' and the ``lowest'' points of the Reeb graph
correspond to  extremal points of $\mathcal{H}_n$.
This means that ```boundary vertices'' typically correspond to
closed trajectories such that, consequently, the wings of the Landau band
are filled by semi-classically localized states.
On the other hand, the ``middle region'' of the Reeb graph contains
loops or nodes which correspond to the open trajectories or
separatrices such that the central part of the Landau band
consists of extended states.

\begin{figure}
\includegraphics[width=80mm]{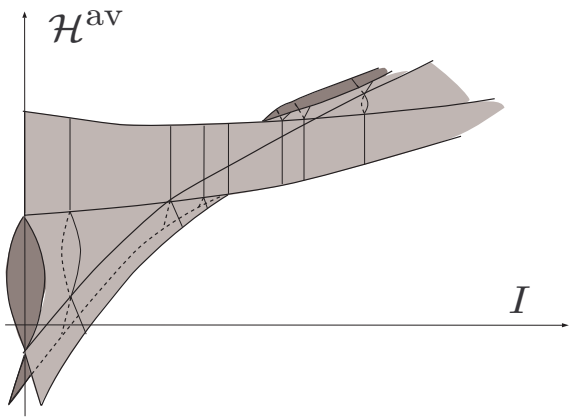}
\caption{The Reeb surface for
the potential $v(X_1,X_2)=\cos(X_1)+\cos(X_2)+2\cos(X_1+X_2)$}\label{fig2}
\end{figure}

\begin{figure}
\includegraphics[width=80mm]{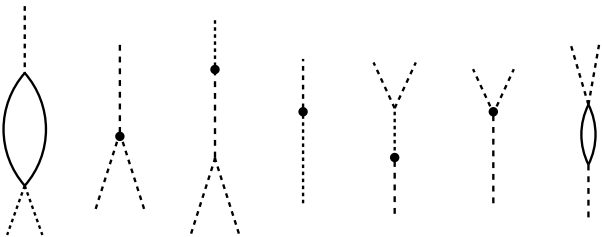}
\caption{The structure of different Landau bands
for the potential $v(X_1,X_2)=\cos(X_1)+\cos(X_2)+2\cos(X_1+X_2)$}\label{fig3}
\end{figure}

Using the previous considerations one can describe
the spectrum globally. To do this, one has to construct
the Reeb graph of ${\mathcal H}^0(I,Y_1,Y_2,\varepsilon_V)$
for each $I$; as $I$ runs through $[0,+\infty)$,
the Reeb graph is transformed and traces out a certain surface
(the \emph{Reeb surface} of the potential $v$). 
We illustrate this by considering the potential
$v(X_1,X_2)=\cos X_1 +\cos X_2
+2\cos(X_1+X_2)$. Using Eq.~(\ref{1}), we get 
\[
\begin{array}{c}
{\mathcal H}^{\rm av}(I,Y_1,Y_2)=I+\varepsilon_V\big(J_0(\sqrt{2I})(\cos Y_1+\cos Y_2)\\
{}+2J_0(\sqrt{4I})\cos(Y_1+Y_2)\big).
\end{array}
\]
Use first $\mathcal{H}^{\rm av}$ instead of $\mathcal{H}^0$,
then the Reeb surface has the form shown in Fig.~\ref{fig2}.
Quantizing $I$ by the Bohr--Sommerfeld rule
$I=E_n$ we select
a discrete subset of the Reeb graphs, which can be obtained
by cutting the Reeb surface by the plane $I=E_n$.
For different indices $n$
we obtain different types of Reeb graphs,
some of which are shown in Fig.~\ref{fig3}.

It is apparent that in all these cases the Landau band
has the asserted structure. For some $n$, the domain
of infinite motion may be very small and tend to a point
(these points are emphasized in Fig.~2), but it always exists.
The inclusion of the correction $\mathcal{H}^0-{\mathcal H}^{\rm av}$ may lead
to the appearance of new vertices and new edges in the Reeb graph near
the branching points; but these edges have length $O(\varepsilon^2_V)$
and hence do not change the asserted spectral structure.
Clearly, spectral regions, corresponding to different Landau bands
may intersect.

To summarize,
we have proposed a semiclassical
explanation of the geometric structure of the spectrum
for the two-dimensional Landau Hamiltonian with
a two-periodic electric field.
Applying an iterative averaging procedure
we approximately, with any degree of accuracy,
separate variables and reduce the original spectral problem
to an infinite family of one-dimensional
eigenvalues problems for Harper-like operators;
each of these operators is the quantization
of a classical Hamiltonian having a torus as the phase space.
Such a reduction provides a convenient tool for 
describing the geometric structure of each Landau band.
Namely, the space of trajectories of each Hamiltonian
on the torus is represented by the so-called Reeb graph.
The projection of the Bohr-Sommerfeld quantization of the graph
onto the energy axis gives the corresponding Landau band.
Clearly, open edges of the graph
represent contractible trajectories;
quantization of these trajectories leads to localized
states in the Landau band. Therefore, the band wings are filled
by the localized states. On the contrary,
open trajectories are always represented by closed
edges, which lie in the middle of the graph;
they generate the extended states near the middle
of the Landau band.

It should be particularly emphasized that different Landau bands
are described, generally speaking, by topologically
different graphs and, therefore, can have different structure.
Such a distinction  of the geometric structure
of the different Landau band
was established numerically in~\cite{PGea} for
various kinds of potentials.

In carrying out this work we had useful discussions with S.~Albeverio,
J.~Avron, E.~D.~Belokolos, V.~S.~Buslaev, P.~Exner,
M.~V.~Karasev, E.~L.~Korotyaev, V.~A.~Margulis, A.~I.~Neishtadt,
L.~A.~Pastur, M.~A.~Poteryakhin, A.~I.~Shafarevich,
P.~Yuditskii, and J.~Zak. To all these people we express a gratitude.

The work is partially supported by the grants DFG-RAS no.~436 RUS 113/572
and INTAS no.~00-257.

\end{document}